\documentclass[journal]{IEEEtran}
\usepackage{xcolor}

\usepackage{cite}

\usepackage{graphicx}
\usepackage{caption}
\usepackage{bm}
\usepackage{amsmath}

\usepackage{stfloats}
\usepackage[english]{babel}
\usepackage{blindtext,tikz}
\usetikzlibrary{calc}

\begin{document}
%
\title{ \fontsize{20}{12}\selectfont Power System Anomaly Detection and Classification Utilizing\\\vspace{0.2cm} WLS-EKF State Estimation and Machine Learning}
%
%
%

\author{Sajjad~Asefi, 
        Mile~Mitrovic, 
         Dragan~Ćetenović, 
         Victor~Levi,
        Elena~Gryazina,
        Vladimir~Terzija
\thanks{Sajjad Asefi is with Department of Electrical Power Engineering and Mechatronics, Tallinn University of Technology, Tallinn, Estonia and also with Center for Energy Science and Technology, Skolkovo Institute of Science and Technology, Moscow, Russia (email: sajjad.asefi@taltech.ee)}
\thanks{Mile Mitrovic, Elena Gryazina and Vladimir Terzija are with Center for Energy Science and Technology, Skolkovo Institute of Science and Technology, Moscow, Russia (email: Mile.Mitrovic@skoltech.ru; E.Gryazina@skoltech.ru; V.Terzija@skoltech.ru)}
\thanks{Dragan Ćetenović and Victor Levi are with the Department of Electrical and Electronic Engineering, University of Manchester, Manchester, UK (e-mail: dragan.cetenovic@manchester.ac.uk; victor.levi@manchester.ac.uk)}
}

\maketitle

\begin{abstract}
Power system state estimation is being faced with different types of anomalies. These might include bad data caused by gross measurement errors or communication system failures. Sudden changes in load or generation can be considered as anomaly depending on the implemented state estimation method. Additionally, considering power grid as a cyber physical system, state estimation becomes vulnerable to false data injection attacks. The existing  methods for anomaly classification cannot accurately classify (discriminate between) the above mentioned three types of anomalies, especially when it comes to discrimination between sudden load changes and false data injection attacks. This paper presents a new algorithm for detecting anomaly presence, classifying the anomaly type and identifying the origin of the anomaly, i.e., measurements that contain gross errors in case of bad data, or buses associated with loads experiencing a sudden change, or state variables targeted by false data injection attack. The algorithm combines analytical and machine learning (ML) approaches. The first stage exploits an analytical approach to detect anomaly presence by combining $\chi^2$-test and anomaly detection index. 
The second stage utilizes ML for classification of anomaly type and identification of its origin, with particular reference to discrimination between sudden load changes and false data injection attacks. The proposed ML based method is trained to be independent of the network configuration which eliminates retraining of the algorithm after network topology changes. The results obtained by implementing the proposed algorithm on IEEE 14 bus test system demonstrate the accuracy and effectiveness of the proposed algorithm.  
\end{abstract}


%
\IEEEpeerreviewmaketitle

\section{Introduction}
\begin{figure*}[!t]
    \centering
    \captionsetup{justification=centering,font=footnotesize}
    \includegraphics[scale = 0.7]{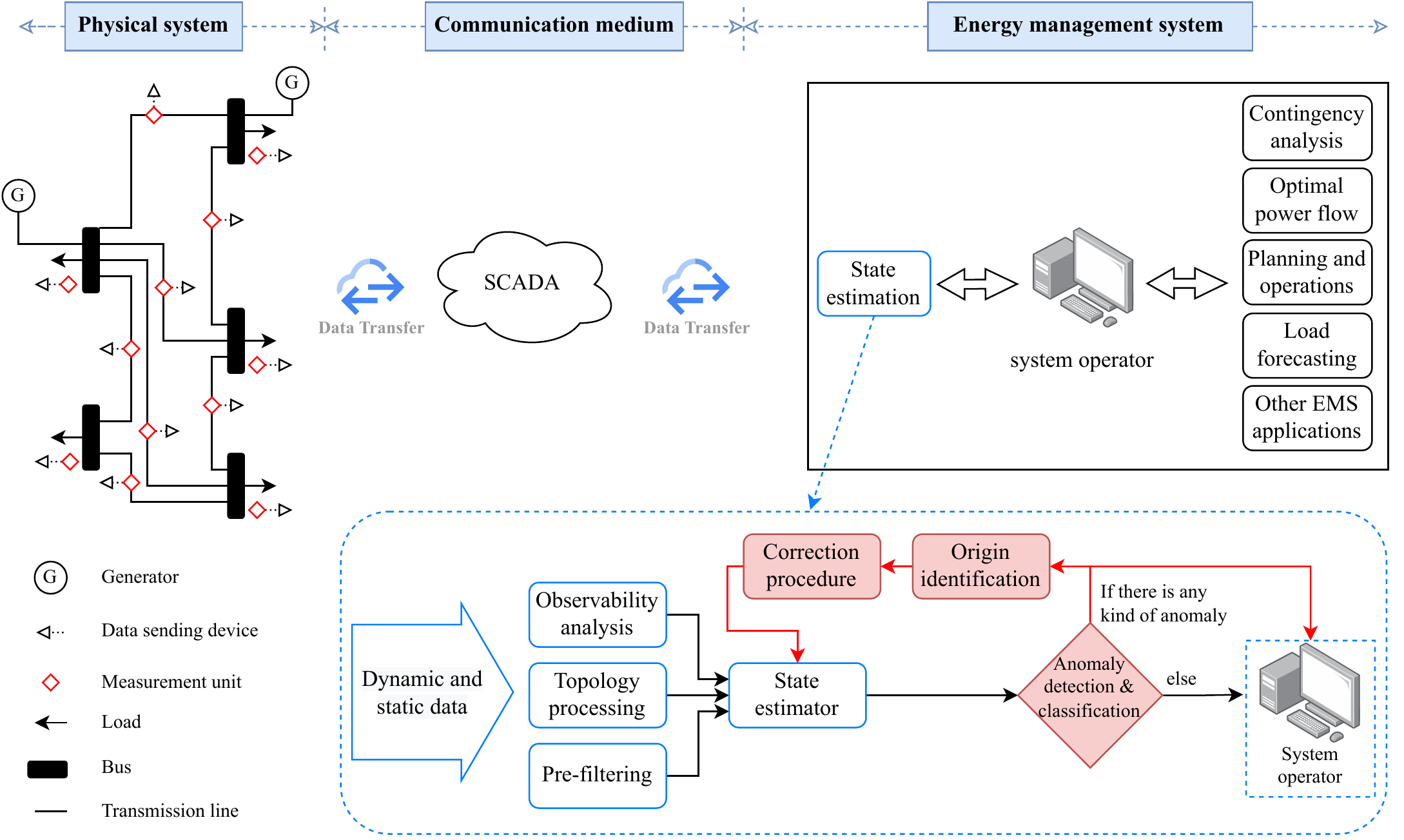}
    \caption{State estimator's role in the power system and position of anomaly detection, classification and identification unit within the state estimator}
    \label{fig: overallSys}
\end{figure*}

\IEEEPARstart{s}{tate} estimation (SE) is the core component of the energy management system (EMS) for power grids. SE provides the most likely values of the voltage magnitudes and phase angles for all buses in the power system. The accuracy of these values is essential for achieving optimal and secure operation of the system \cite{gomez2011taxonomy}. Fig. \ref{fig: overallSys} demonstrates the connection between physical, communication and energy management systems, emphasizing the role of anomaly detection, classification and identification units within the state estimator.

Power system SE can be subjected to different types of anomalies that might spoil the accuracy of the estimated states. Inter alia, these can be anomalies like bad data (BD), sudden change in bus injections or false data injection attack (FDIA). BD is caused by unexpected errors in sensors or communication medium. Besides, network model parameters might contain BD. Sudden change in bus injections can be either sudden load change (SLC) or sudden generation change (SGC), depending on whether consumer or generator is connected to the bus. Severe SLC is usually caused by serious variations in industry load or by disconnection/reconnection of a large portion of the load. Although SLC was under the scope of many research work in the past, SGC emerges as a new challenge since penetration of uncertain renewable energy sources is increasing incessantly  \cite{nishiya1982dynamic, valverde2010unscented}. Both SLC and SGC will further lead to the sudden change in the system operation point. FDIA is a type of prefect interacting BD \cite{zhao2018generalized}. FDIA is amongst the most hazardous cyberattacks which targets data integrity. It has attracted industry and research community’s attention recently \cite{nejabatkhah2021cyber, reda2021taxonomy, mukherjee2022novel,hock2020using}. Regardless of the anomaly type, it is essential to detect the anomaly presence as soon as it occurs, as well as to classify (discriminate between) different types of anomalies and identify their origin in order to enable proper countermeasures against each of them in the correction phase (see Fig. \ref{fig: overallSys}). 

Techniques for detection and classification of anomalies depend on which SE method is implemented. Assuming that normalized measurements’ residuals follow a standard Gaussian distribution, their sum of squares are expected to follow a $\chi^2$ distribution \cite{gomez2018electric, abur2004power}. This resulted in $\chi^2$-test to be widely used as BD detection method in conventional weighted least square (WLS) power system static state estimation (SSE). Besides, largest normalized residual (LNR) test has shown good detecting capabilities but also the ability to identify measurement(s) corrupted with bad data \cite{abur2004power}. 
LNR test found its role in bad data detection stage for hybrid Voltage Source Converter - HVDC/AC transmission systems, where it has been improved by integrating the Gaussian mixture model algorithm \cite{ayiad2021state}. LNR method is also used to cope with bad data in multi-energy applications \cite{zang2019robust}. Other approaches suitable for WLS framework are recently studied to mitigate the impact of outliers, like Hampel’s redescending and the Schweppe–Huber generalized M-estimators \cite{ho2019robust}. Unlike WLS, forecasting-aided state estimation (FASE) based on Kalman filtering utilizes process model yielding state forecasting. FASE combines state predictions with observed measurements which facilitates BD detection and classification. To detect BD in phasor measurement units (PMUs), generalized maximum-likelihood unscented Kalman Filter (UKF) applies projection statistics to a two dimensional matrix of measurement innovations and predicted states considering strong temporal correlations between them \cite{zhao2017robust}. A fault tolerant second-order extended Kalman Filter (EKF) based on discrete-time nonlinear Luenberger-type observer has been presented in \cite{wang2019second} to mitigate the adverse effects of BD. 


Considering WLS SSE employs only current snapshot of measurements, $\chi^2$-test conducted over measurement residuals will not recognize the presence of the SLC. On the other hand, due to existence of the process model, SLC will affect FASE performance. This might be used to enable SLC detection. In \cite{nishiya1982dynamic, valverde2010unscented}, normalized measurement innovations are tested against predefined threshold to detect anomaly presence, while skewness of distributions and $\chi^2$-test of goodness of fit of normalized innovations are proposed to discriminate between BD and SLC. In \cite{da1983state}, same detection technique has been used whilst the logical check routine is applied for classification purpose. In \cite{dang2020robust}, an UKF based on minimum error entropy is presented to avoid abnormalities such as SLC. A maximum-correntropy-based  EKF estimator is presented in \cite{massignan2020tracking} for discrimination between BD and SLC, incorporating both supervisory control and data acquisition (SCADA) and PMU measurements. In \cite{ma2019unscented}, an UKF with generalized correntropy loss is introduced to suppresses the effect of outliers by utilizing inverse of the exponential function of innovations for update of measurements noise covariance matrix. 

In the presence of FDIA, previously discussed detection and classification methods might fail with a high probability. If FDIA is stealthy designed, measurement residuals of WLS estimator will not change. This makes $\chi^2$-test, conducted over measurement residuals, blind to this kind of anomaly and thus FDIA stays undetected. Detection problem can be solved by taking advantage of FASE through measurement innovations or by combining the results of WLS SSE and FASE. In \cite{vzivkovic2018detection}, it has been shown that a metric defined as weighted difference between WLS and FASE estimates is able to detect presence of FDIA if system operates in quasi steady state. However, effects of SLC and network topology changes on detection performance are not considered in \cite{vzivkovic2018detection}. In \cite{pei2021deviation}, FDIA is detected by applying an exponential function of innovations. Although the case of FDIA targeting multiple states at the same time (hereinafter, multi-state FDIA) is studied in \cite{pei2021deviation}, the case when load is abruptly changed at multiple buses simultaneously (hereinafter, multi-bus SLC)
as well as the effects of topology changes on FDIA detection are not taken into account. Due to the fact that stealthy FDIA and SLC will affect the statistics of the FASE outputs in a similar way, discrimination between FDIA and SLC is hard to be achieved even using FASE. However, in \cite{nath2019quickest} the authors have considered the statistical properties of innovation vector to distinguish between SLC and FDIA; the objective function of the optimization problem aims to minimize the detection delay taking into account an upper bound for false alarm rate. However, the later reference did not consider cases with multi-state FDIA or multi-bus SLC, neither how changes in the network topology affect detection and classification performance.

As discussed above, determining an analytical approach for detection and classification of the discussed three anomalies is a difficult task. Nowadays, similar to other research areas, application of machine learning (ML) methods in power system is increasing due to low mathematical dependency on system model and self-learning capability of these methods \cite{xie2020review}. Additionally, proper accuracy and effectiveness makes ML based methods a suitable choice for detection and classification of anomalies \cite{sayghe2020evasion}. Apart from supervised ML methods commonly utilized in the literature, such as logistic regression, k-near neighborhood, random forest and/or extreme gradient boosting, combination of artificial neural networks and ML, so called deep learning, is also broadly applied in the area of power systems.    
In \cite{mestav2019learning}, a deep neural network architecture that integrates a universal BD detection technique using a binary hypothesis testing scheme has been presented. A matrix completion approach is proposed for SE of distribution networks in \cite{liu2019robust}, aiming to minimize the weighted sum of the measurement residuals to suppress the effects of BD. Bayesian BD detection method within a deep learning based SE scheme is used in \cite{mestav2019bayesian}.
An ensemble correlation based detector with adaptive statistics, presented in \cite{nagaraj2020ensemble}, compares squared Mahalanobis distance of new measurement samples with an adaptive threshold in order to detect and classify FDIA. Authors in \cite{esmalifalak2017detecting} have introduced two ML based algorithms for FDIA detection: if the measurements are labeled (i.e. normal and tampered measurements are specified), detection is performed using support vector machine algorithm; in the case of unlabeled measurements, detection is done utilizing statistical characteristics of historical measurements. A deep learning based method utilizing a feed-forward artificial neural network has been implemented for FDIA detection in  \cite{ashrafuzzaman2018detecting}. 
In \cite{mestav2019learning,liu2019robust,mestav2019bayesian,nagaraj2020ensemble,esmalifalak2017detecting, ashrafuzzaman2018detecting}, detection and classification of SLC and multi-state FDIA, as well as analyzing the effects of network topology change on detection and classification of FDIA and SLC, have not been discussed. In another words, the algorithms in the mentioned literature are not trained to deal with such anomalies. 

To overcome the previously discussed shortcomings of the existing methods, in this paper we developed a new algorithm to reliably detect and classify power system anomalies, namely BD, SLC and FDIA, with particular reference to discrimination between SLC and FDIA that are not happening simultaneously. The proposed algorithm consists of two stages. At the first stage, the anomalies which are undetectable using conventional detection technique (i.e. $\chi^2$-test), are detected by analytical approach based on  comparative analysis of WLS SSE and EKF FASE estimation results. Second stage utilizes ML based algorithm for classification of anomaly type and origin. Four supervised ML algorithms have been tested: logistic regression, k-near neighborhood, random forest and extreme gradient boosting. These algorithms have been tested individually and their performances are compared against each other. Besides reliable classification of anomalies, our work also aims to increase the robustness of ML algorithm against changes in the network configuration. This eliminates the need to retrain the algorithm once network topology changes. The focus of this research work is on anomaly detection, classification and identification of its origin, while the countermeasures are not under scope of this paper. 

To the best of the authors’ knowledge, this is the first attempt to discriminate between single/multi–bus SLC and single/multi–state FDIA considering different network configurations and different load levels. The novelties of the research work can be stated as follows:
\begin{itemize}
    \item New algorithm for detection and classification of single/multi–bus SLC and single/multi–state FDIA is developed. Developed algorithm utilizes anomaly detection index to enable detection, whilst supervised ML is used for classification of mentioned anomalies. 
    
    \item New solution for classification of FDIA and SLC is proposed. Proposed solution eliminates the adverse effects of topology changes by utilizing the features associated only with the buses, such as nodal measurements, normalized measurement innovations, estimated and predicted values of measurements and system states.
    
    
    
    \item ML algorithm training speed is accelerated by applying an optimal feature selection method, i.e. maximum relevance -- minimum redundancy.

\end{itemize}

The rest of the paper has been organized in the following manner: Section \ref{sec: sysDescript} introduces power system model, basics of static and forecasting aided state estimation and preliminary considerations for BD, SLC and stealthy FDIA; Section \ref{sec: method} describes the proposed methodology, that is the combination of FASE and WLS SSE followed by the details of the supervised ML based methods and their evaluation metric, as well as the optimal feature selection method; In Section \ref{sec: results-discussion}, various scenarios of SLC and FDIA occurrence are presented and the numerical results of the simulations are discussed; Paper is concluded by Section \ref{sec: concl}.

\section{Problem formulation}
\label{sec: sysDescript}

This section introduces state transition and measurement models followed by EKF FASE and WLS SSE algorithms in brief. Initial deliberations about BD, SLC and FDIA are also provided.

\subsection{State transition and measurement model}
Taking into account the slow enough changes in the system operating point,
state transition model can be described by linear stochastic equation \cite{cetenovic2018optimal, valverde2010unscented, zhao2019power}:

\begin{equation}
    \bm \hat{x}_{t} = \bm A_{t-1}\bm \hat{x}_{t-1} + \bm g_{t-1} + \bm \omega_{t-1}
\label{eq: state_trans}
\end{equation}
where $\bm \hat{x}$ is the state vector composed of $n = 2N-1$ elements (bus voltages and phase angles except slack bus’s phase angle), $t$ is the sampling time, $\bm A$ is the state transition matrix, $\bm g$ is the trend vector, $\bm \omega$ is process noise which has a Gaussian distribution with zero mean and covariance matrix $\bm Q$, and $N$ is the number of buses.
The widely used approach for updating matrix $\bm A$ and vector $\bm g$ is Holt's exponential smoothing regression \cite{da1983state}.


Set of $m$ measurements considered in this paper is composed of active and reactive power flows, active and reactive power injections, and voltage magnitudes. Relation between these measurements and the states at time $t$ can be expressed as follows:

\begin{equation}
        \bm z_t = \bm h(\bm \hat{x}_t) + \bm e_t.
    \label{eq: meas_model}
\end{equation}
where $\bm z$ is measurement vector, $\bm h$ is set of non-linear equations, and $\bm e$ represents vector of measurement noise assumed to be Gaussian distributed with zero mean and covariance matrix $\bm R$. 

\subsection{Extended Kalman Filter based Forecasting Aided State Estimation}
FASE is a special application of dynamic state estimation concept in which the dynamics of the states are negligible. FASE utilizes both state transition and measurement model represented in \eqref{eq: state_trans} and \eqref{eq: meas_model}, respectively.
Kalman filtering has commonly been used as the optimal solution for many data tracking and forecasting tasks \cite{basetti2016power}.
The Kalman Filter is an estimator that uses previous state and current observation to estimate the current state.

Considering the power system, the measurement function is non-linear in nature. If the system is non-linear, an extension of classical Kalman Filter, so called EKF can be utilized through the linearization of the non-linear model via Taylor series \cite{jin2021improved}. There are also other extensions of Kalman Filter to deal with non-linearity of the system, such as UKF \cite{valverde2010unscented}, Particle Filter \cite{del1997nonlinear}, iterated EKF  \cite{bretas1989iterative}, Ensamble Kalman Filter \cite{houtekamer1998data}, Second order Kalman Filter \cite{gelb1974applied}, Cubature Kalman Filter \cite{basetti2022square}, to name a few. However, for the sake of simplicity and less computation burden, a first order EKF based estimator has been utilized in this study.

\subsubsection{Prediction equations}
$\hat{\bm x}_{t-1}$ and $\hat{\bm P}_{t-1}$ are estimated state vector and estimated states' error covariance matrix at time step $t-1$, respectively. Assuming that $\hat{\bm x}_{t-1}$ and $\hat{\bm P}_{t-1}$ are known, the states of the system can be predicted using following equations \cite{cetenovic2018optimal}:
\begin{IEEEeqnarray}{l}
        \tilde{\bm x}_{t} = \bm A_{t-1}\hat{\bm x}_{t-1} + \bm g_{t-1}  \label{eq: predic1}\\
        \tilde{\bm P}_{t} = \bm A_{t-1}\hat{\bm P}_{t-1}\bm A^T_{t-1} + \textit{\textbf{Q}}_{t-1}   \label{eq: predic2}
\end{IEEEeqnarray}
where $\tilde{\bm x}_{t}$ is predicted state vector and $\tilde{\bm P}_{t}$ indicates the state prediction error covariance matrix at time step $t$.

\subsubsection{Filtering equations}
After predicting the states using \eqref{eq: predic1} and \eqref{eq: predic2}, and receiving a new set of measurements $\bm z_t$ at time step $t$, estimated state $\hat{\bm x}_t$ and its covariance matrix $\hat{\bm P}_{t}$ can be obtained as \cite{cetenovic2018optimal}:
\begin{IEEEeqnarray}{l}
       \bm\nu_{t} = \bm z_t - \bm h(\tilde{\bm x}_t)  \label{eq: est1}\\
        \bm S_{t} = \bm H_{t}\tilde{\bm P}_{t} \bm H^T_{t} + \bm R_{t}   \label{eq: est2}\\
        \bm K_{t} = \tilde{\bm P}_{t} \bm H^T_{t} \bm S^{-1}_{t}      \label{eq: est3}\\
        \hat{\bm x}_{t} = \tilde{\bm x}_{t} + \bm K_{t} \bm\nu_{t}      \label{eq: est4}\\
        \hat{\bm P}_{t} = \tilde{\bm P}_{t} - \bm K_{t} \bm S_{t} \bm K^T_{t} \label{eq: est5}
\end{IEEEeqnarray}
where $\bm H$ represents Jacobian of $\bm h$ that is evaluated at $\tilde{\bm x}_{t}$; $\bm \nu$, $\bm S$ and $\bm K$ indicate innovation vector, innovation covariance matrix and Kalman gain, respectively.

\subsection{Weighted Least Squares based Static State Estimation}
The SSE results from a further simplification in which the state transition information is completely disregarded and only the nonlinear measurement function \eqref{eq: meas_model} is kept. As a result, SSE has no recollection of the states at earlier time steps and unlike FASE, it lacks the capacity to track system states transition \cite{zhao2019power}. Furthermore, SSE implies that the state vector can only be observed using the most recent set of available measurements.

Based on the concept of maximum likelihood and assuming the measurement errors are independent and follow a standard Gaussian distribution \cite{abur2004power, gomez2018electric}, SSE problem can be written in the following format: 
\begin{equation}
 min \text{ } [\bm z - \bm h(\bm \hat{x})]^T \times \bm R^{-1} \times [\bm z- \bm h(\bm \hat{x})]
\label{eq: SSE_objective_function}
\end{equation}

The solution for the optimization problem in \eqref{eq: SSE_objective_function} is called WLS that can be solved using Newton-Raphson iterative process \cite{gomez2018electric}. 

\subsection{Bad data detection and identification}
If the measurement residuals follow Gaussian distribution, their sum of squares would have a $\chi^2$ distribution. Combination of $\chi^2$-test and LNR is widely used within the WLS SSE framework for bad data detection (BDD) and identification of BD's origin. Its steps are given as follows  \cite{gomez2018electric}:

\begin{itemize}
  \item Calculate the following objective function after solving the SE problem:
  \begin{equation}
      J_{BDD}(\hat{\bm x}) = \sum_{i=1}^{m} \frac{[\bm z_i - \bm h_i(\hat{\bm x})]^2}{\sigma_i^2}
      \label{eq: Chi_Sq_BDD}
  \end{equation}
    where $\sigma_i$ is the standard deviation of the $i$-th measurement ($\sigma^2_i$ corresponds to the $i$-th diagonal element of matrix $\bm R$).
  \item From $\chi^2$ distribution table pick up the value corresponding to probability $p$ and ($m - n$) degrees of freedom.
  \item If $J_{BDD}(\hat{\bm x}) \geq \chi_{(m-n),p}^2$ holds, then there is high probability of existence of a BD, otherwise most probably there is no BD.
  \item If BD is detected, calculate normalized residual for each measurement, $\bm r^{Norm}_i$:
\begin{equation}
    \bm r^{Norm}_i
= \frac{|\bm z_i - \bm h_i(\hat{\bm x})|}{\sqrt{\bm{\Omega}_{ii}}}
      \label{eq: LNR}
\end{equation}
where $\bm \Omega$ is the residual covariance matrix.
\item If $\bm r^{Norm}_k$ is the largest normalized residual and $\bm r^{Norm}_k > \tau$ ($\tau = 3$), then $k$-th measurement will be suspected as BD.
\end{itemize}

\subsection{Sudden load/generation change}
One of the events that might change the power system state abruptly is SLC or SGC. This might happen due to failure of different power system components, such as circuit breakers or generation units. Considering constant increase in penetration from renewable energy sources into the power system, their intermittent nature might be another reason for sudden state change. It is to be noted that $\chi^2$-test carried out over measurements’ residuals obtained via WLS SSE is unable to detect SLC/SGC. On the other hand, application of FASE can be helpful for detection of these events due to advantages that state transition model brings. For the sake of brevity, in this research we have focused on SLC and modeled it as a load shedding at different buses but we point out that similar considerations can be applied to the case of SGC as well. More details about SLC modeling is provided in Section \ref{sec: results-discussion}.

\subsection{False data injection attack}
With the expansion of the power system and application of various communication mediums, the possibility for cyber-attacks has increased. Considering the power system as cyber-physical system, its data must have three main factors, so called availability, integrity and confidentiality. Two well-known attacks in the power system are FDIA and denial of service (DoS), which threaten integrity and availability of the data, respectively \cite{mohan2020comprehensive}. Although the target medium for both types of attack is the communication medium, FDIA can lead to critical issues to the secure and economic operations in the power system if it evades being detected by conventional BDD. FDIA can mislead the system operator that the system operates in a normal and secure state, while in reality it is not. Also, the operator may be persuaded to take expensive and unnecessary corrective measures like load shedding or rescheduling generator units. Conventional BDD, such as $\chi^2$-test, utilizes measurement residuals, while stealthy FDIA endeavors to keep residuals unchanged. This might lead to conventional BDD failure against stealthy FDIA. 

Assuming the adversarial has perfect knowledge of the system and receives the same data as the system operator, he can be capable of manipulating the measurements in a way that BDD will be bypassed. The attack vector $\bm a$ is of the same size as the measurement vector but with non-zero elements ($\bm a_i$) correspondent only to measurements under the attack. So, under the attack, the $i$-th measurement will have the following model:
\begin{equation}
    \bm z_i^{a}= 
\begin{cases}
    \bm z_i + \bm a_i   & \text{if $i$-th measurement is attacked} \\
    \bm z_i   & \text{otherwise}
\end{cases}
\label{eq: attack_vec}
\end{equation}
 
 From practical point of view, \eqref{eq: attack_vec} infers that the adversarial needs to have access to measurements and the ability to manipulate their values to the values determined by the attack vector. 
 
 Without losing generality, let us assume that the adversarial has obtained the same measurement and state values as the system operator. The FDIA can bypass the BDD if $\bm a_i = \bm h(\bm x_i^a) - \bm h(\bm \hat{x}_i)$, where $\bm x_i^a = \bm \hat{x}_i + \bm c_i$ and $\bm c_i$ represents the change adversarial would like to make to the $i$-th state. Taking into account that generally $J_{BDD}(\hat{\bm x}) \leq \chi_{(m-n),p}^2$ holds in the absence of anomaly, then  (for simplicity, index $i$ has been omitted) \cite{zhao2015forecasting}:
\begin{equation}
\begin{split}
\bm r^a & = \bm z^a - \bm h(\bm x^a) \\
        & = \bm z + \bm a - \bm h(\bm \hat{x} + \bm c) \\
        & = \bm z + \bm a - \bm h(\bm \hat{x} + \bm c) + \bm h(\bm \hat{x}) - \bm h(\bm \hat{x}) \\
        & = \bm z - \bm h(\bm \hat{x}) + \bm a - [\bm h(\bm \hat{x} + \bm c) - \bm h(\bm \hat{x}) ] \\
\end{split}
\label{eq: Norm_Res_BDD}
\end{equation}

The above equation proves the feasibility of stealthy attack in case the adversarial has enough information regarding the network measurements and parameters to build $\bm a$ in a way that $a = \bm h(\bm \hat{x} + \bm c) - \bm h(\bm \hat{x})$. It is worth noting that in this research the conventional BDD is assumed to be $\chi^2$ method, as explained above. More details of modeling FDIA is presented in Section \ref{sec: results-discussion}.

\section{Proposed algorithm for anomaly detection, classification, and identification}
\label{sec: method}

\begin{figure}[!h]
    \centering
    \captionsetup{justification=centering,font=footnotesize}
    \includegraphics[width = \columnwidth]{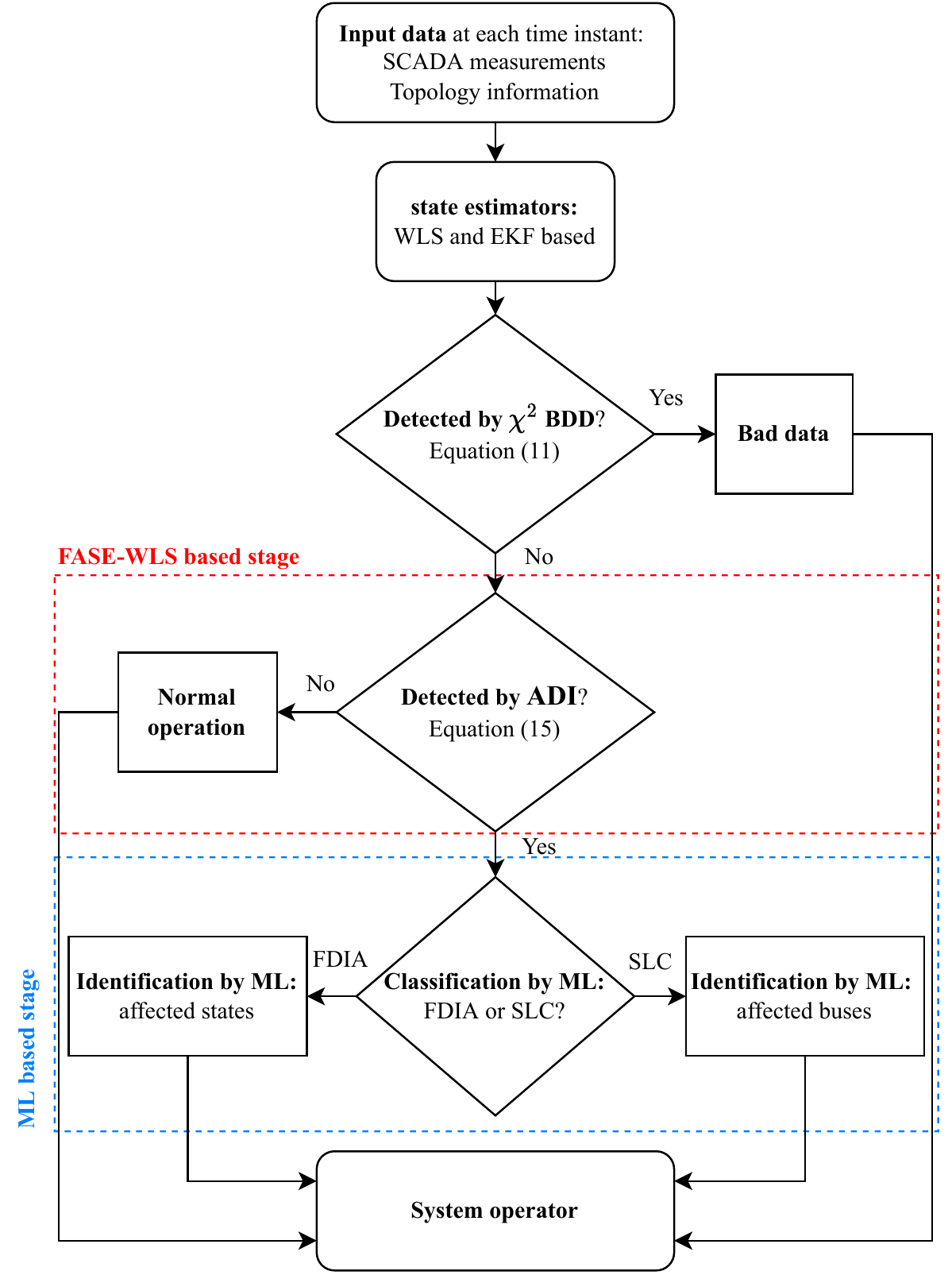}
    \caption{Flowchart demonstration of the proposed algorithm for anomaly detection and classification}
    \label{fig: Analytical_ML_comb_stages}
\end{figure}

This Section describes the proposed method’s workflow. As it is shown in Fig. \ref{fig: Analytical_ML_comb_stages}, the workflow  combines analytical and ML approaches which are explained in the following subsections in detail. Input data, i.e. observed SCADA measurements and status of switching devices, is being delivered to the EMS and after that SE is performed using WLS and EKF. Measurement residuals obtained at WLS output are used to carry out $\chi^2$-test in order to check for BD. If some measurements are corrupted with BD, $\chi^2$-test will rise the flag indicating BD presence. If SLC or FDIA occurs, $\chi^2$-test will not detect anomaly presence and the process will continue to FASE-WLS based stage. In the FASE-WLS based stage, the estimated states that are obtained by WLS and EKF are utilized to form an anomaly detection index \eqref{eq: BDD_EKFbased}. In case the index value is equal or higher than the specific threshold, anomaly is detected; otherwise, system is considered to be in the normal operation mode. Although the index is capable to detect presence of those anomalies for which $\chi^2$-test stays blind, it is still not capable to classify the anomaly type due to similar impacts which SLC and FDIA have on WLS and EKF estimates. To classify detected anomaly as SLC or FDIA, workflow leverages the ML based stage. After classification of the anomaly, the next step in the ML based stage is to identify at which bus(es) SLC has happened, or which state(s) has been affected by FDIA.

\subsection{FASE-WLS based approach}

In this paper, in order to detect SLC or FDIA presence, an anomaly detection index (ADI) that combines WLS and EKF estimates is employed \cite{vzivkovic2018detection}:

\begin{equation}
    ADI_i = \frac{|\hat{x}_i^{WLS} - \hat{x}_i^{EKF}|}{\sqrt{p_{ii}}}\hspace{5 mm}i=1, 2, ..., 2n-1
    \label{eq: BDD_EKFbased}
\end{equation}
where $\hat{x}_i^{WLS}$ \textit{i}-th state variable estimated by the WLS; $\hat{x}_i^{EKF}$ \textit{i}-th state variable estimated by the EKF; $p_{ii}$ is the \textit{i}-th diagonal element of EKF estimated states' error covariance matrix, $\hat{\bm P}$. 
If $\displaystyle\max_{i}\{ADI_i\} \geq \gamma$, SLC or FDIA presence is detected. Here, $\gamma$ represents detection threshold that has to be selected to clearly discriminate between normal operation and anomalies like SLC and FDIA. BD is less relevant for the threshold setting because, when it occurs, it is expected to be detected by $\chi^2$-test. In this paper, EKF has been selected as a Kalman Filter extension for non-linear systems. However, the anomaly detection index (15) can be equaly utilized for anomaly detection if any other type of Kalman Filter has been used. 

\subsection{Supervised machine learning algorithms}
As previously mentioned, SLC and FDIA can have similar impact on WLS and EKF outputs. This fact makes it hard to classify anomaly as SLC or FDIA using ADI or similar indices dependent on the system model. Because of low dependency to the system model and using only historical data, application of ML algorithm is an appropriate choice for SLC and FDIA classification. The supervised tree-based algorithms for classification have been employed to classify SLC or FDIA, as well as to determine the buses (or states) that have been affected by SLC (or FDIA). Accordingly, the Random Forest (RF) and Extreme Gradient Boosting (XGB) algorithms were applied. Other ML algorithms such as Logistic Regression (LR) and K-Near Neighbors (KNN) were also considered and compared.  


Classification is a ML supervised concept which categorizes a set of data into classes. Classification algorithms determine the class of new samples based on past samples during training. Depending on the number of classes, the problem can be considered as a binary or multi-class classification. In our case, the classification of FDIA or SLC is a binary classification task, while determination of the state or bus that has been affected by anomaly represents a multi-class classification problem.

Proposed supervised algorithms work with labeled data. The number of labeled samples used in this paper depends on the task which we consider. Each collected sample is represented by $\bm x$ and $y$ coordinates $\{(\bm x_i, y_i), i=1,...m_{total}\}$, where $\bm x$ and $y$ describes input and output of the models, respectively; $m_{total}$ is total number of samples. Each input is a $n_x$-dimensional feature vector $\bm x_i \in R^{n_x}$. However, each output is a class labeled that represent FDIA or SLC $[y_{i}\in \{0,1\}]$, or states/buses of considered IEEE 14-bus system. 

To train and test models, data set samples were splitted in training $\{(\bm x_i, y_i), i=1,...m_{train}\}$ and testing $\{(\bm x_i, y_i), i=1,...m_{test}\}$ subsets; $m_{train}$ and $m_{test}$ are training and testing number of samples, respectively. Splitting was conducted in a stratified fashion, containing approximately the same percentage of samples of each labeled class. 80\% of data samples were used for training and 20\% data samples were tested. The controllable hyperparameters are tuned using sequential optimization with gradient boosting \cite{friedman2002stochastic} as a surrogate probability model of the objective function.

\subsubsection{RF} It is an non-linear ensemble classifier based on multiple decision trees, using a randomly selected subset of training samples and variables. RF is designed using the bootstrap aggregating, also known as the bagging technique based on the combination of weak decision tree algorithms in parallel that will improve the effectiveness of the prediction \cite{breiman2001random}.  

The RF architecture represents a collection of $t_{RF}$ randomized classification trees with tree-structured classifier $h_{RF}$. The RF is expressed as:
\begin{equation}
     h_{RF}(\bm x, \bm 
    \eta_j), ~ j=1,...,t_{RF}
\end{equation}
where $\bm \eta_j$ are the  independent identically distributed random vectors.

$h_{RF}$ trees are created in parallel, independent of one another, using the bootstrapped data sets for growing the trees. Random subset of variables at each step of tree growth is used to split the node. The quality of the splitting is measured by applying the gini impurity criterion G:

\begin{equation}
    \mathcal{G} = \sum_{i=1}^{c} p(i)\cdot(1-p(i))  
\end{equation}
where $p(i)$ is the probability of samples belonging to class $i$ at a given node.

The final RF prediction $ \hat{y}$ is the majority vote over collection of trees:
\begin{equation}
    \hat{y} =  ~ majority ~ vote ~ \{h_{RF}^j(\cdot)\}_{j=1}^{t_{RF}}
\end{equation}

For more details reader can refer to \cite{breiman2001random}. 

\subsubsection{XGB} It is designed using boosting technique and attempts to build a robust model from the number of weak tree classifiers in series. XGB is a specific implementation of the Gradient Boosting method, which uses more accurate approximations by employing second-order gradients and advanced regularization \cite{chen2016xgboost}.

For a given number of $t_{XGB}$ sequentially connected decision tree models $h_{XGB}$, the XGB can be expressed as: 
\begin{equation}
    \hat{y} = \sum_{j=1}^{t_{XGB}} h^j_{XGB}(\bm x) 
\end{equation}
where $h^j_{XGB}(x) = w^j_q(x)$. $w^j_q$ is the score of the corresponding leaf $q$ in the $j$-th tree.  

To optimize the functions used in the model, XGB minimizes the following regularized loss function $\mathcal{L}$:
\begin{equation}
    \mathcal{L}(y, \hat{y}) = \sum_{i=1}^{m_{train}} l(y_{i},\hat{y}_i) + \sum_{j=1}^{t_{XGB}}\Omega(h^j_{XGB})
\end{equation}
where $l(\cdot)$ is a logistic loss function. The second term $\Omega$ of the loss function is the regularization term which penalizes the complexity of the model:
\begin{equation}
    \Omega(h_{XGB})  = \gamma T + \frac{1}{2}\lambda 
    \sum_{q=1}^{T} w^2_q
\end{equation}
where $\gamma$ is the pseudo-regularization
hyperparameter and $T$ is the number of leaves in the tree.

Since XGB ensemble model includes functions as parameters and cannot be optimized using conventional optimization methods, the training of the model is executed in manner of $t$ steps: 
\begin{equation}
    \mathcal{L}(t) = \sum_{i=1}^{m_{train}} l(y_{i},\hat{y}^{t-1}_i + h^t_{XGB}(x_i)) + \Omega(h^t_{XGB})
\end{equation}
where $h^t_{XGB}$ is greedily added to improve model. For more details reader can refer to \cite{chen2016xgboost}. 

\subsection{Evaluation metric}
The \textit{macro} F1-score metric is used to evaluate the accuracy of the FDIA or SLC classification, as well as to evaluate the accuracy of the anomaly's origin identification \cite{opitz2019macro}.
F1-score is a \textit{harmonic} mean of the \textit{precision} $Pr$ and \textit{recall} $Re$. For a particular predicted output vector $\hat{y}$ and ground-truth $y$, the percentage of F1-score is computed as follows:
\begin{equation}
    F1 = 2 \cdot \frac{Pr \cdot Re}{Pr + Re} \times 100
    \label{eq: F1_evalMetric}
\end{equation}
The \textit{precision} represents the number of \textit{True Positives (TP)} over \textit{TP} plus the number of \textit{False Positives (FP)}:
\begin{equation}
    Pr = \frac{TP}{TP+FP}
\end{equation}
The \textit{recall} is the \textit{TP} over the \textit{TP} plus the number of \textit{False Negatives (FN)}:
\begin{equation}
    Re = \frac{TP}{TP+FN}
\end{equation}
As an example for FDIA and SLC classification, \textit{TP} is the number of correctly determined FDIAs. \textit{FP} is the number of SLCs  determined as FDIAs, and \textit{FN} is the number of FDIAs determined as SLCs. The similar logic is applicable for multi-bus SLC and multi-state FDIA.


The macro F1-score is calculated as the arithmetic mean over the F1-scores of each class:
\begin{equation}
    macro~F1 = \frac{1}{c} \sum_{i=1}^{c} F1_i .
\end{equation}

\subsection{Maximum Relevance -- Minimum Redundancy}

Increase in the size of the system will lead to increase in the number of features which will consequently increase optimization complexity of the ML algorithm. Maximum Relevance Minimum Redundancy (MRMR) is a feature selection algorithm for finding the minimal-optimal subset of features \cite{ding2005minimum, zhao2019maximum}. Minimal-optimal methods select a small set of features that have the maximum possible predictive power by eliminating irrelevant features. Accordingly, the model optimization complexity is reduced.

The purpose of this method is to reduce the number of input features that linearly increases with the system size:
\begin{equation}
    n_x = \alpha  N - \beta
\end{equation}
where $\alpha = 16$ that is sum of features for each bus and $\beta = 10$ is sum of features related to slack bus. Accordingly, by employing MRMR we are able to select just a few main features to achieve the high enough level of accuracy.

MRMR works iteratively. At each iteration $i$, it identifies the best feature $f_b$ that has maximum relevance with respect to the target variable and minimum redundancy with respect to the features that have been selected at previous iterations. Since nonlinear dependency exists between input and target variables, to compute maximum relevance of the feature $f_b$, we use mutual information method \cite{smith2015mutual}. In contrast, for computing minimum redundancy of the feature $f_b$, we employ Spearman's rank based correlation \cite{gauthier2001detecting} because of handling non-normality data.

The score for each feature $f_b$ at each iteration $i$ is computed as follows \cite{zhao2019maximum}:
\begin{equation}
    score_i(f_b) = \frac{relevance(f_b|target)}{redundancy(f_b| {f_b}_{selected ~ until ~ i-1})}
\end{equation}
where the best feature $f_b$ at iteration $i$ is the one having the highest score.

In this paper, MRMR has been applied to select a few main features that will result in reducing the optimization complexity during training of the ML algorithms. As a parameter of optimization complexity, training time has been considered. The reduced training times are also presented in the next section for the utilized supervised ML algorithms.

\subsection{Solution for topology changes}
As mentioned before, the transmission system faces topology changes (i.e. change in the network configuration). This will require ML algorithm to be retrained. Retraining ML algorithm can be time consuming and inefficient depending on the size of the system. To eliminate the need for retraining the ML algorithm after the change in network topology, features related to the power lines are excluded and only features associated with the buses are applied for ML algorithm training. These features are the ones associated only with the buses such as: a) nodal measurements and normalized measurement innovations of voltage magnitude and active/reactive power injection; b) estimates and predictions of voltage magnitude, phase angle and active/reactive power injection. Five topologies are used to train and afterwards examine the accuracy of anomaly classification and identification of its origin.  These five topolgies include original topology of the IEEE 14 bus system \cite{christiepower} and four new topologies obtained by disconnecting an existing line from the original topology and making new connection. These new topologies are specified in Table \ref{tab: topologyChange}.  
 
\begin{table}[h!]
    \centering
    \captionsetup{justification=centering,font=footnotesize}
    \caption{Connection and disconnection of lines for topology changes}
    \begin{tabular}{|c|c|c|}
    \hline
        Topology & Disconnected line &  Connected line\\
        number  & From bus -- To bus &  From bus -- To bus\\
        \hline
            1 & 5\phantom{1} -- \phantom{1}6\phantom{1} & 1\phantom{1} -- \phantom{1}6\phantom{1} \\
         \hline
            2 & 6\phantom{1} -- \phantom{1}13 & 6\phantom{1} -- \phantom{1}14 \\
         \hline
            3 & 4\phantom{1} -- \phantom{1}9\phantom{1} & 4\phantom{1} -- \phantom{1}10 \\
         \hline
            4 & 2\phantom{1} -- \phantom{1}4\phantom{1} & 3\phantom{1} -- \phantom{1}5\phantom{1} \\
         \hline
    \end{tabular}
    \label{tab: topologyChange}
\end{table}

The obtained data considering these togologies are utilized for training and testing of the ML algorithms and the results are presented in the next section.

\section{Results and discussions}
\label{sec: results-discussion}
In this section, various possible scenarios of single/multi-bus SLC and single/multi-bus FDIA occurrence are considered for analysing the accuracy of the proposed methodology.

The consecutive optimal power flows were run to get true values of states and measurements over the time. 
A noise having Gaussian distribution with zero mean and 0.01 standard deviation is added to true measurements to get the observed measurements. To model a BD, corresponding measurement is corrupted with a random error which does not fall under the predefined Gaussian distribution. For modeling SLC, specified amount of the load is curtailed at the desired time instant during the execution of the consecutive optimal power flow. In the case of FDIA, the observed measurements are modified according to the attack vector. State estimation is carried out under normal and abnormal operation, and afterwards the proposed algorithm for anomaly detection, classification and identification of its origin is executed.

The overall data set contains numerous SLC and FDIA scenarios. For SLC, following scenarios have been considered: \begin{itemize}
    \item Single-bus SLC is simulated for different buses.
    \item For each bus, single-bus SLC is simulated numerous times; every next time, different portion of the load has been curtailed from the corresponding bus.
    \item Multi-bus SLC is simulated for different combinations of buses.
    \item For each combination of buses, multi-bus SLC is simulated numerous times; every next time, different portions of loads at corresponding buses are curtailed.
    \item All the above mentioned events are simulated considering different topologies.  
\end{itemize} 
For FDIA, following scenarios have been considered:
\begin{itemize}
    \item Single-state FDIA is simulated for different state variables in the system.
    \item For each state variable under the attack, single-state FDIA is simulated numerous times; every next time, the corresponding element in the attack vector has a different value.
    \item Multi-state FDIA is simulated for different combinations of state variables.
    \item For each combination of state variables under the attack, multi-state FDIA is simulated numerous times; every next time, corresponding elements in the attack vector have different values.
    \item Again, all the above mentioned events are simulated considering different topologies.  
\end{itemize}

The number of the state variables that can be affected by the FDIA depends on how many measurements can be accessed by the adversarial. It is hardly feasible for the adversarial to access all the measurements in the system especially in the case of large scale power systems. It is more realistic that the adversarial can access and manipulate the measurements within a local area.
In this paper, it has been assumed that the adversarial is capable of manipulating state variables associated with the maximum $4$ buses simultaneously. Moreover, FDIA can target both voltage magnitude and phase angle (both in single or multi-state attack scenarios), while in the reported literature researchers mostly have focused on FDIA on voltage magnitudes. 

As mentioned in section \ref{sec: method} and shown in Fig. \ref{fig: Analytical_ML_comb_stages}, the second stage of the classification process is based on ML algorithm. Supervised ML algorithms' performance are compared and the results are presented. Presented results are given for the case of SLC and FDIA classification and identification of their origin. In case BD occurs, $\chi^2$-test will detect the anomaly and recognize it as BD, while LNR will be sufficient to identify the measurements corrupted with the BD.

ML algorithms are developed in Python using scikit-learn and scikit-optimize libraries. They are tested on Intel Core i7-5500U CPU @ 2.40GHz and 8GB of RAM. Fig. \ref{fig: ClassificationSingleAnomalyAllTopologies} to Fig. \ref{fig: multiStateFDIAidentification} illustrate the performance of the supervised ML algorithms for anomaly classification and identification of its origin. Each figure demonstrates \textit{macro} F1-score and training time without utilizing MRMR method (specified as "WO MRMR" in the figures). Additionally, the results for the case considering the features selected by MRMR method are presented for each ML algorithm (specified as "MRMR" in the figures). Total number of features is $214$ and the number of features selected by MRMR method is given for each case study.

\subsection{Detection of SLC and FDI}
\begin{figure}[!h]
    \centering
    \captionsetup{justification=centering,font=footnotesize}
    \includegraphics[trim={0.7cm .2cm 0.7cm 0},width = \columnwidth]{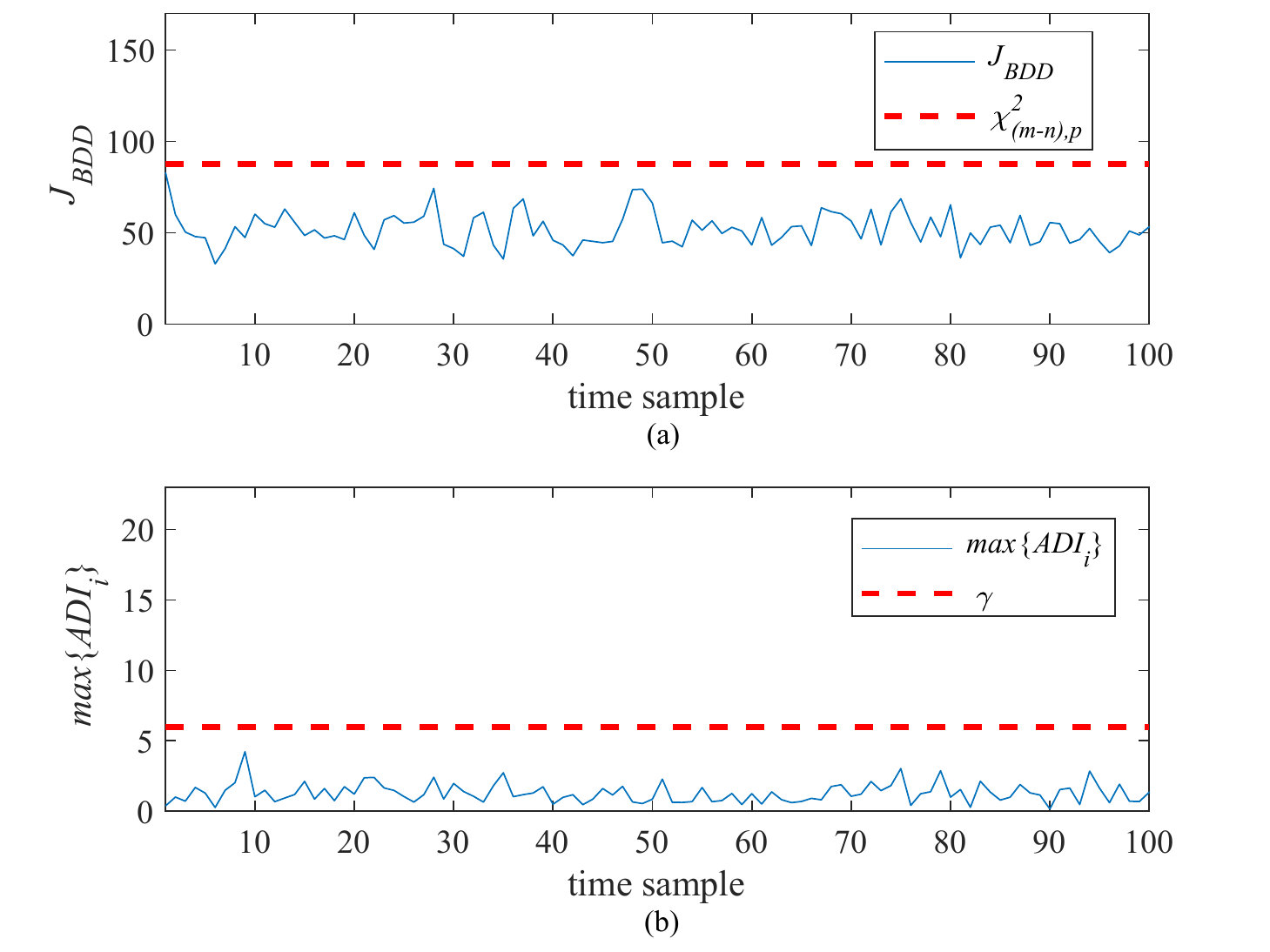}
    \caption{Detection index value when the system is under normal operation: (a) $\chi^2$-test (b) ADI}
    \label{fig: analytic_Normal}
\end{figure}

\begin{figure}[!h]
    \centering
    \captionsetup{justification=centering,font=footnotesize}
    \includegraphics[trim={0cm .3cm 0cm 0}, width = \columnwidth]{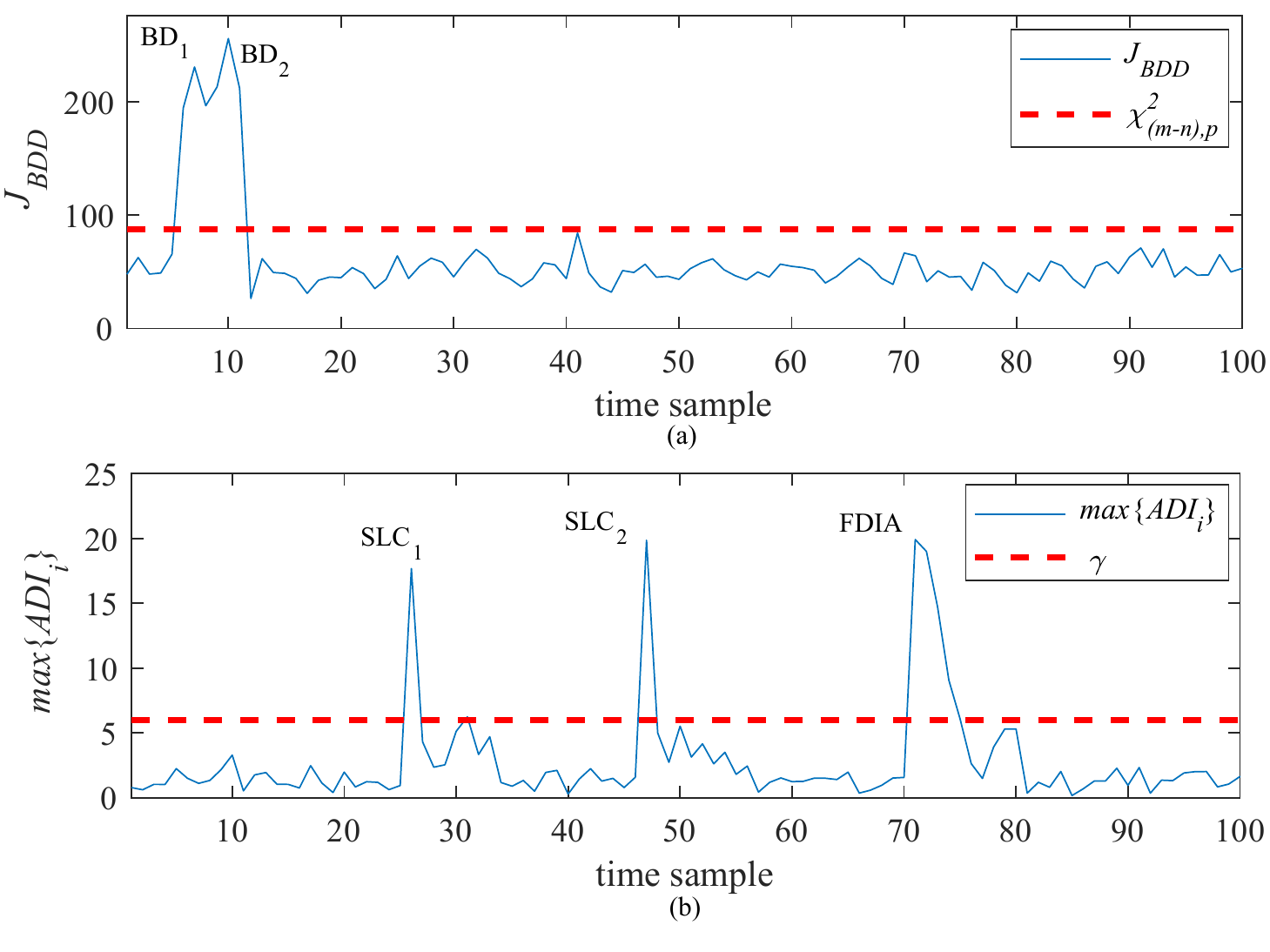}
    \caption{Detection index value when the system is affected by anomaly: (a) $\chi^2$-test (b) ADI}
    \label{fig: analytic_event}
\end{figure}

An example of anomaly detection by the proposed algorithm is illustrated here. Fig. \ref{fig: analytic_Normal} illustrates detection indices value when the system is under normal operation, while Fig. \ref{fig: analytic_event} demonstrates the case when BD and single bus/state SLC/FDIA happen in the system. 

In Fig. \ref{fig: analytic_Normal}, load at every bus is assumed to linearly decrease during the simulation period from $100\%$ to $95\%$ of its nominal value, making slow changes in the system state. When system is under normal operation, value of the index should be below the specified thresholds. 
In order to increase (or decrease) the sensitivity for anomaly detection, the threshold  can be lowered (or raised). However, if the threshold is set too low, this may increase the number of false alarms. On the other hand, if the threshold is set too high, anomaly presence might not be detected. To properly select the threshold for ADI, $\gamma$, it is necessary to run simulations under normal and abnormal operation conditions for each particular test system. In the case of IEEE 14 bus system, extensive simulations have shown that  $\gamma = 6$ can clearly distinguish between normal operation and anomalies like SLC and FDIA.

Fig. \ref{fig: analytic_event} shows the ADI and $\chi^2$-test values when the system is affected by anomaly. BD, SLC and FDIA are happening at bus $14$ but at different time instants (not simultaneously). $BD_1$ is related to BD, when power injection measurement value at bus $14$ contains $5\%$ error at $t = 5$, and $BD_2$ at $t = 10$ refers to the situation when this error has been removed. $SLC_1$ represents $20\%$ load shedding at bus $14$ which happens at $t = 6$. $SLC_2$ corresponds to the situation when the load at bus $14$ is restored at $t = 46$. FDIA tends to increase the voltage magnitude at bus $14$ for $0.05$ p.u. starting from $t = 71$ and persist till the end of simulation. ADI also might be applicable for higher error values of BD, but $\chi^2$-test is resilient enough for this anomaly. It is obvious that ADI is highly capable of detecting SLC and FDIA anomalies, while they bypass the conventional BDD. Yet, this method is not able to classify the occurred anomaly as SLC or FDIA. This is the main concern within this research and application of ML algorithm is proposed to address it.

\subsection{Classification and identification of single bus/state SLC/FDIA}
In this case study, the performance of the ML algorithms for classification of single bus SLC and single state FDIA are presented. After classification of the anomaly, the ML algorithms are utilized to identify the origin of the SLC or FDIA. As stated before, these anomalies are happening at different time samples (not simultaneously).

\begin{figure}[!h]
    \centering
    \captionsetup{justification=centering,font=footnotesize}
    \includegraphics[width = \columnwidth, height=5.5cm]{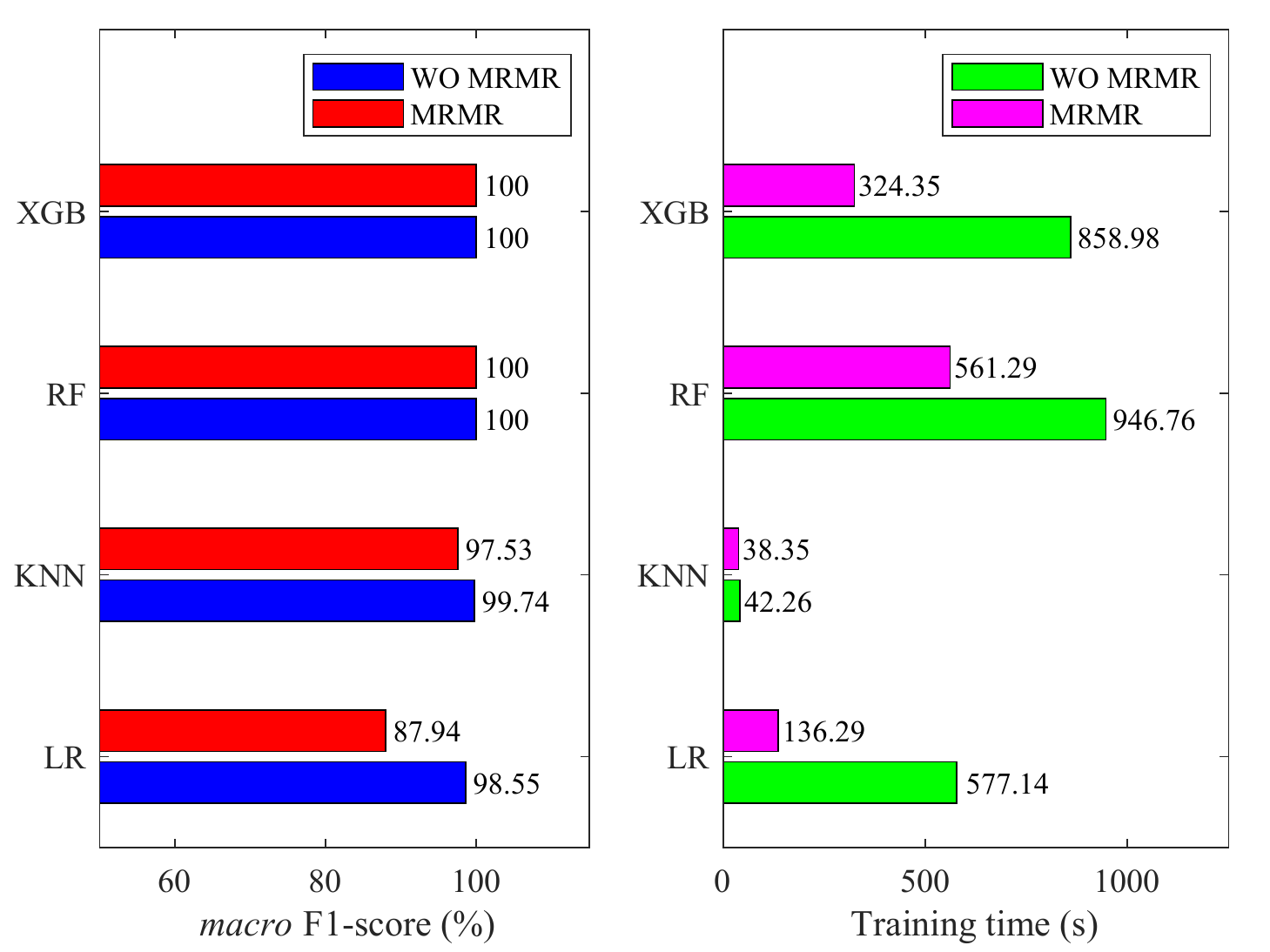}
    \caption{Single bus/state SLC/FDIA classification using all topologies to gather training and testing data}
    \label{fig: ClassificationSingleAnomalyAllTopologies}
\end{figure}

As indicated in the Fig. \ref{fig: ClassificationSingleAnomalyAllTopologies}, if all features are utilized, the classification accuracy of each method is higher than 98\%, which can be considered as acceptable. Due to the fact that some of the features might be redundant (superfluous) or less relevant for training of the ML algorithms, MRMR has been applied to select the most relevant features. In the case of single bus/state SLC/FDIA classification using all topologies to gather training and testing data, the number of features selected by MRMR is $70$. This has helped to reduce the training time of the ML algorithms. Although the accuracy of LR and KNN methods slightly decreases, in the case of RF and XGB methods the accuracy remains the same.

\begin{figure}
    \centering
    \captionsetup{justification=centering,font=footnotesize}
    \includegraphics[width = \columnwidth, height=5.5cm]{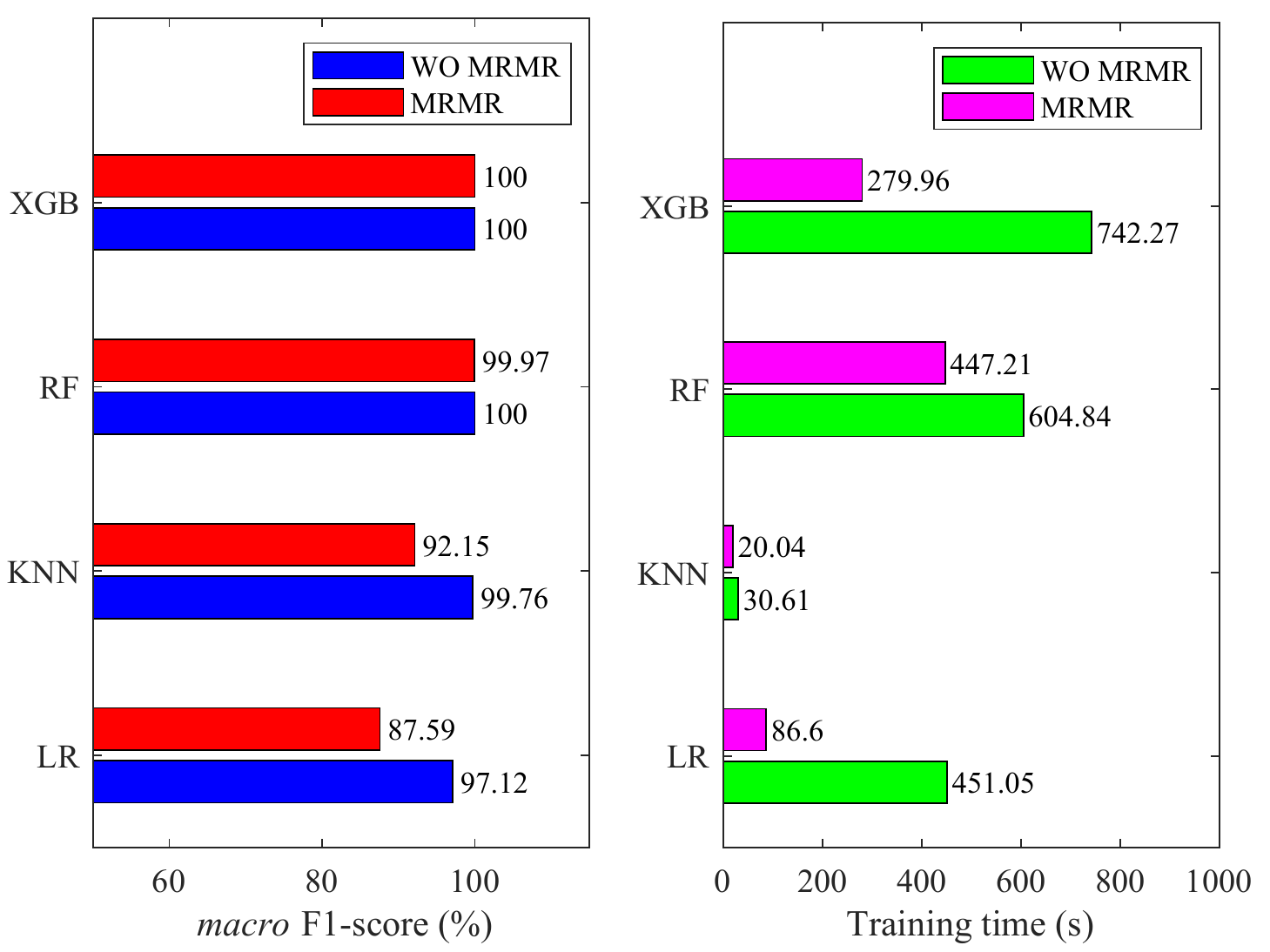}
    \caption{Single bus/state SLC/FDIA classification using untrained topologies to gather testing data}
    \label{fig: ClassificationSingleAnomalySomeTopologies}
\end{figure}

The results given in Fig. \ref{fig: ClassificationSingleAnomalyAllTopologies} correspond to the case when both training and testing data set contains the data obtained under $5$ different topologies. This means that ML algorithms are tested using the same network topologies for which they have been trained. To check how ML algorithms perform against untrained network topologies, $3$ out of $5$ topologies have been used in the training phase of the ML algorithms, while in the testing phase ML algorithms are tested using the data corresponding to the other $2$ topologies. The results are shown in Fig. \ref{fig: ClassificationSingleAnomalySomeTopologies}. 

It is clear that all examined ML algorithms show an acceptable accuracy if they encounter the data corresponding to the topologies which have not been used in their training phase. This means that SLC and FDIA classification can be achieved without retraining the ML algorithm once network topology changes. This is because the proposed methodology excludes the features associated with the power lines and utilizes only the features associated with the buses. If MRMR is used to optimize the number of features, it will turn out that $80$ features would be sufficient. Based on the classification accuracy the algorithms can be sorted in the following descending order: XGB, RF, KNN and LR.

\begin{figure}
    \centering
    \captionsetup{justification=centering,font=footnotesize}
    \includegraphics[width = \columnwidth, height=5.5cm]{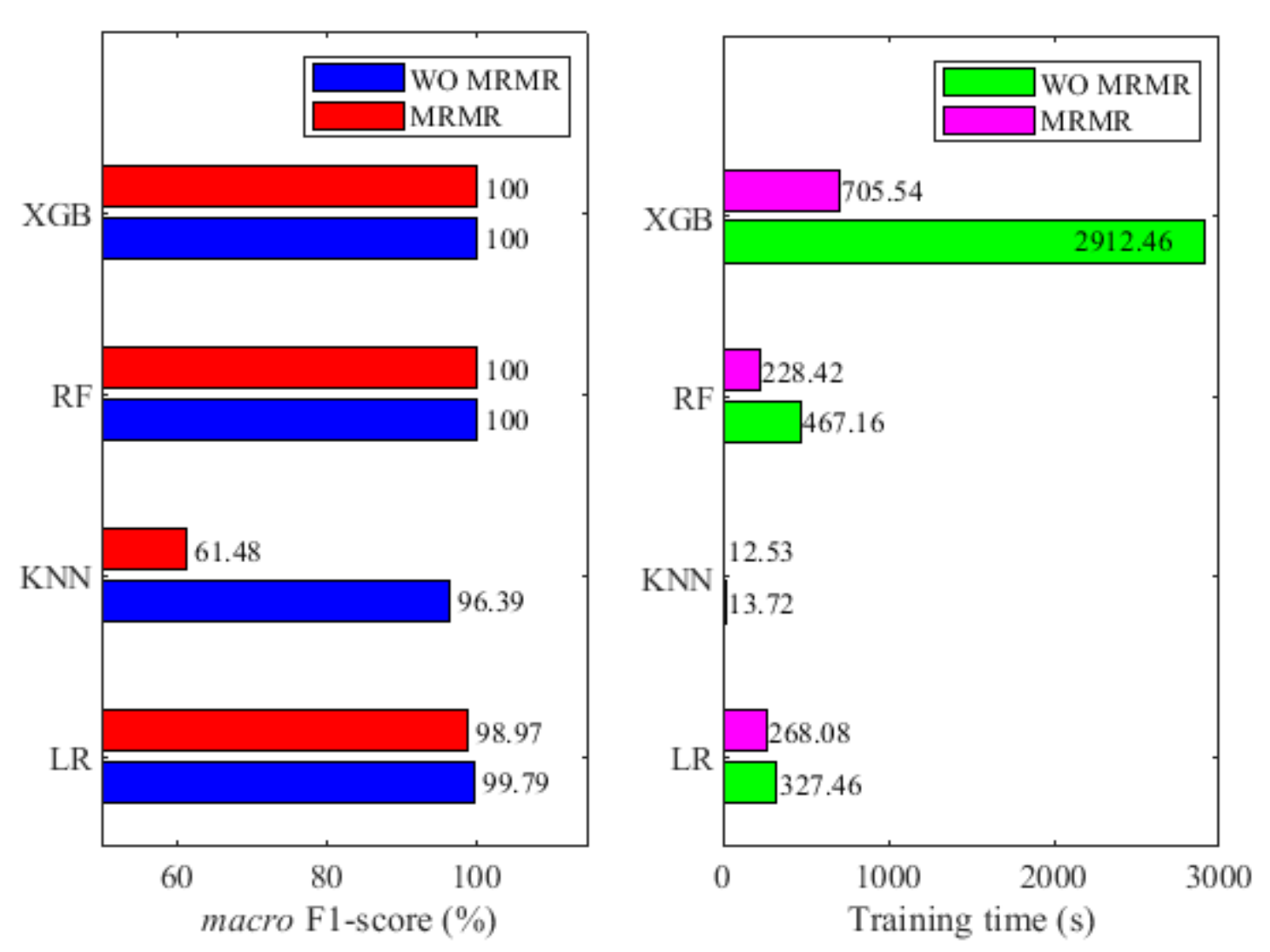}
    \caption{Identification of single bus SLC}
    \label{fig: singleBusSLCidentification}
\end{figure}
\begin{figure}
    \centering
    \captionsetup{justification=centering,font=footnotesize}
    \includegraphics[width = \columnwidth, height=5.5cm]{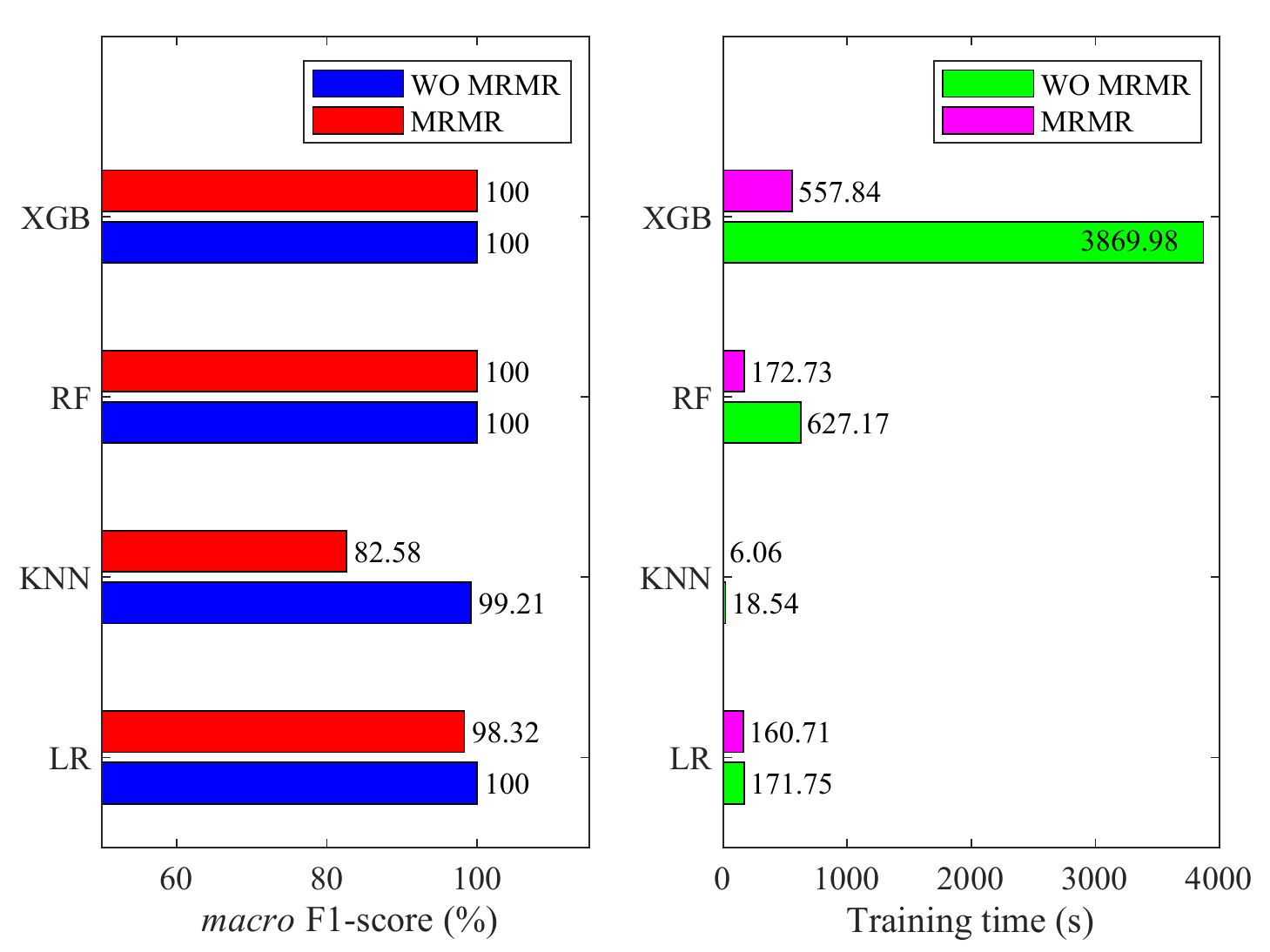}
    \caption{Identification of single state FDIA}
    \label{fig: singleStateFDIAidentification}
\end{figure}

After classification of the anomaly, the bus associated with load experiencing a sudden change, or state variable targeted by FDIA have to be identified. The results for identification of SLC and FDIA origin are presented in Fig. \ref{fig: singleBusSLCidentification} and Fig. \ref{fig: singleStateFDIAidentification}, respectively.

Based on the demonstrated results, it is clear that ML algorithms are successful in identifying the bus (or the state variable) which is affected by SLC (or FDIA). Furthermore, MRMR algorithm provides the optimal number of features for identification of anomaly's origin, which in the case of single bus SLC and single state FDIA is $40$ and $15$, respectively. As in the case of classification, RF and XGB also provide the best identification accuracy, while using MRMR decreases the training time for these two algorithms significantly.

\begin{figure}
    \centering
    \captionsetup{justification=centering,font=footnotesize}
    \includegraphics[width = \columnwidth, height=5.5cm]{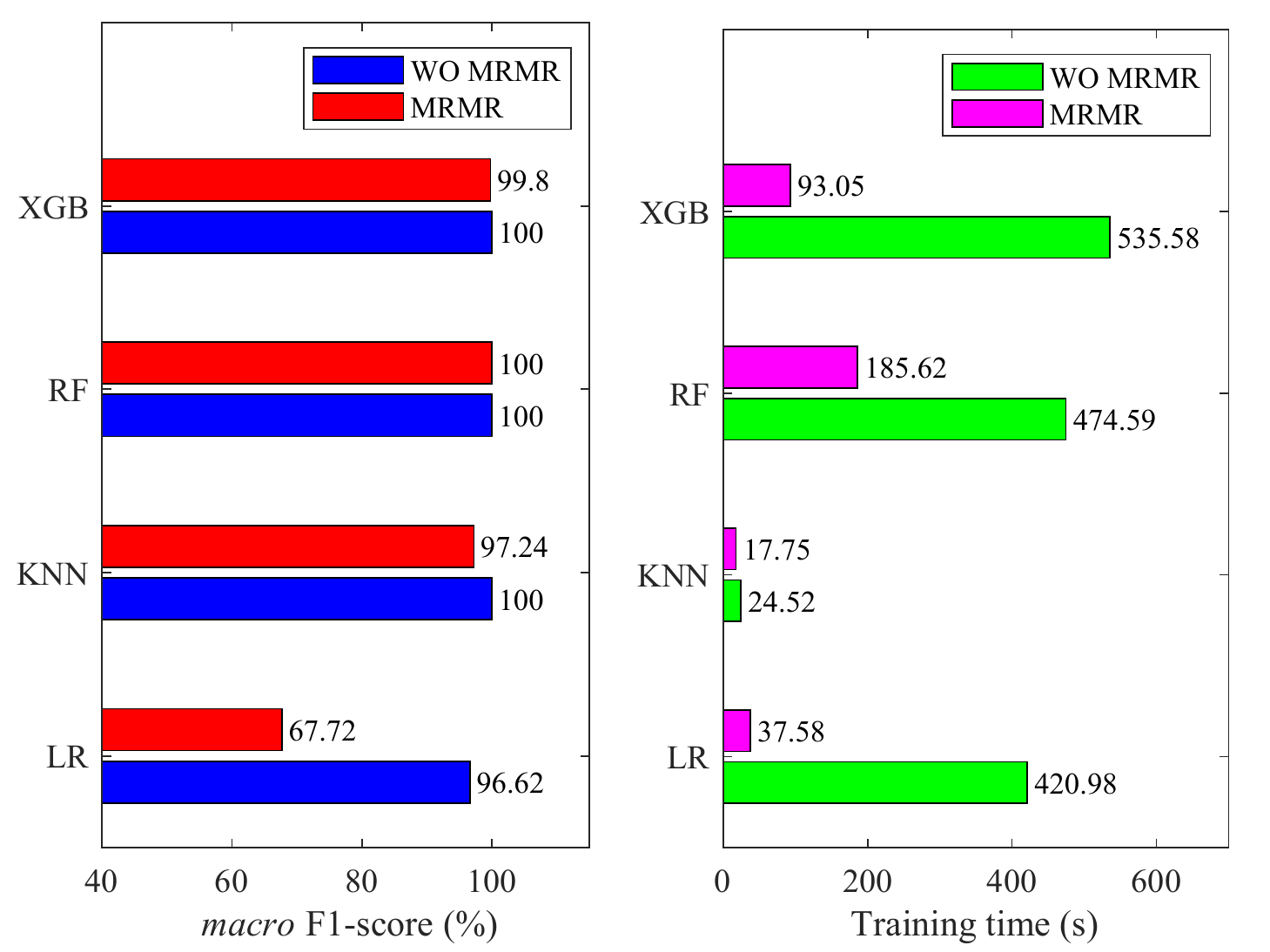}
    \caption{Multi-bus/state SLC/FDIA classification using all topologies to gather training and testing data}
    \label{fig: ClassificationMultiAnomalyAllTopologies} 
\end{figure}

\subsection{Classification and identification of multi bus/state SLC/FDIA}
\begin{figure}
    \centering
    \captionsetup{justification=centering,font=footnotesize}
    \includegraphics[width = \columnwidth, height=5.5cm]{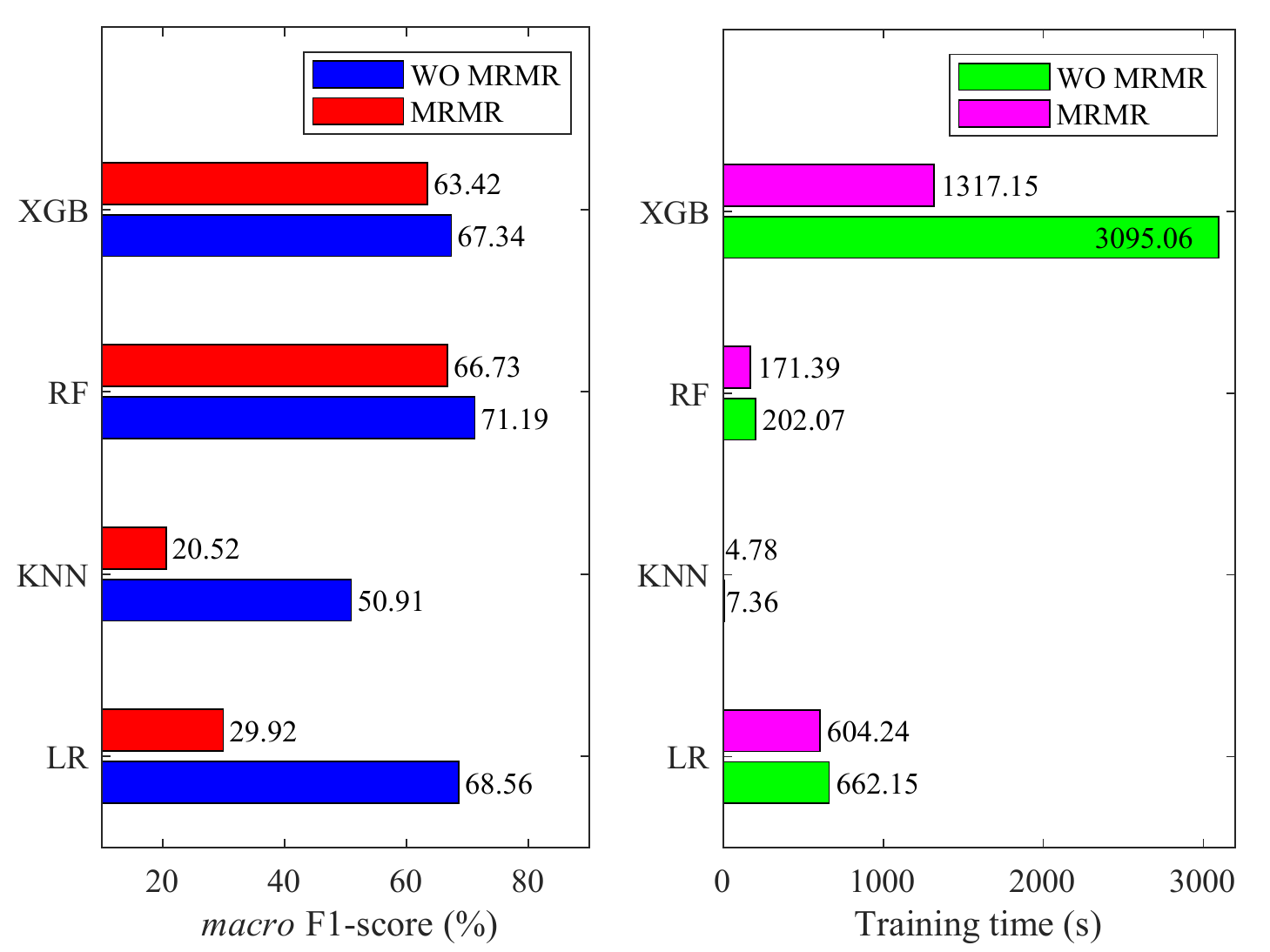}
    \caption{Identification of multi-bus SLC}
    \label{fig: multiBusSLCidentification} 
\end{figure}

\begin{figure}
    \centering
    \captionsetup{justification=centering,font=footnotesize}
    \includegraphics[width = \columnwidth, height=5.5cm]{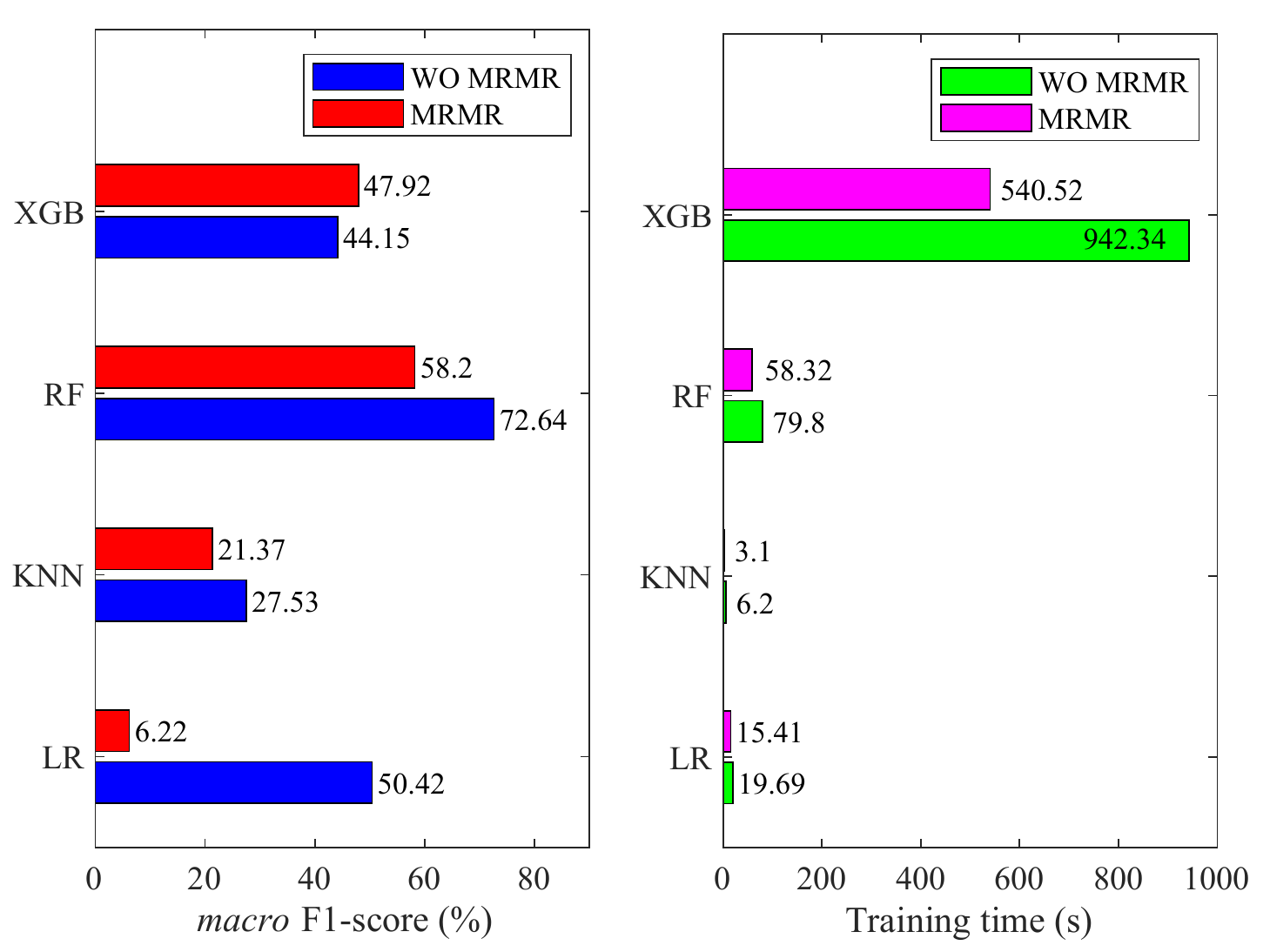}
    \caption{Identification of multi-state FDIA}
    \label{fig: multiStateFDIAidentification} 
\end{figure}

In this case, classification of multi-bus SLC (i.e., SLC is happening at different buses simultaneously) and multi-state FDIA (i.e., multiple states have been targeted by FDIA) is analyzed, as well as identification of the origin of these two kinds of anomalies. 
The results for classification of multi-bus SLC and multi-state FDIA are presented in Fig. \ref{fig: ClassificationMultiAnomalyAllTopologies}.
The results for identification of the buses associated with loads experiencing a sudden change and the results for state variables targeted by FDIA are demonstrated in Fig. \ref{fig: multiBusSLCidentification} and Fig. \ref{fig: multiStateFDIAidentification}, respectively. 
The number of features selected by MRMR algorithm for classification of anomaly is $30$. The number of selected features for identification of multi-bus SLC is $150$, while this number is $70$ for identification of multi-state FDIA.

Based on the results presented in Fig. \ref{fig: ClassificationMultiAnomalyAllTopologies}, it can be concluded that the ML algorithms have a satisfying accuracy for classification of multi-bus SLC and multi-state FDIA. As before, MRMR algorithm helps to reduce the training time. Considering only those features selected by MRMR, RF provides the best classification accuracy followed by XGB, KNN and LR.


The accuracy of the ML algorithms is highly related to the amount of data available for their training. Based on the results, it can be seen  that the amount of data is quite sufficient for classification of multi-bus/state SLC/FDIA. However, for accurate identification of the anomaly's origin, an increased amount of data is required due to the fact that the number of possible combinations of their origin (both for multi-bus SLC and multi-state FDIA) is very huge. 


\section{Conclusion}
\label{sec: concl}



This paper presents a novel solution to detect and classify anomalies such as BD, SLC and FDIA, as well as to identify their origin. Anomalies that bypass the $\chi^2$-test are successfully detected using an anomaly detection index. After that, a ML algorithm is applied to classify anomalies and identify their origin. Based on the obtained results, the proposed algorithm is capable of accurate detection and classification of the anomalies. 

It has been demonstrated that utilizing the features associated only with the buses eliminates the need for retraining the ML algorithm once the network topology changes. Furthermore, application of an optimal feature selection method alleviates the optimization complexity of the ML algorithm. Besides saving the time and the computational resources, these aspects make system operator capable of fast response in case anomaly occurs. 

Detection and classification of different types of anomalies in case of their simultaneous occurrence, along with identification of their origin and designing the suitable countermeasures against them, can be considered as future research directions.

\section*{Acknowledgement}
Work of S.Asefi, M.Mitrovic, E.Gryazina and V.Terzija was supported by Skoltech and The Ministry of Education and Science of Russian Federation, Grant Agreement No 075-10-2021-067, Grant identification code 000000S707521QJX0002.
Work of D.Ćetenović and V.Levi was supported by the Engineering and Physical Sciences Research Council (EPSRC) of UK (Grant No. EP/S00078X/2 and Grant No. EP/T021969/1), and the Ministry of Education, Science and Technological Development of the Republic of Serbia (Grant No. 451-03-68/2022-14/200132 with University of Kragujevac - Faculty of Technical Sciences Čačak). 


%





\bibliographystyle{IEEEtran}
\bibliography{main.bib}

\ifCLASSOPTIONcaptionsoff
  \newpage
\fi

\end{document}